\documentclass[12pt]{article}
\usepackage{a4wide}

\begin{document}

\title{Quantum fields in curved space-time,  semiclassical gravity, quantum gravity phenomenology,  \\ and analogue models \\ {\large Report on session D4 at GR20}} 



\author{Christopher J Fewster$^{1,a}$  and
        Stefano Liberati $^{2,b}$ \\
{\small $^1$Department of Mathematics, 
University of York, Heslington, York, YO10 5DD, UK.}\vspace{2mm}\\
{\small $^2$SISSA/ISAS, Via Bonomea 265, 34136, Trieste, Italy, and} \\
{\small INFN, Sezione di Trieste, Via Valerio, 2, 34127, Trieste, Italy}\\[2mm] 
 {\footnotesize  ~$^a$ \texttt{chris.fewster@york.ac.uk}~,~$^b$ \texttt{liberati@sissa.it} }
}




\maketitle

\begin{abstract}
The talks given in parallel session D4 at the 20th International Conference on General Relativity and Gravitation (Warsaw 2013) are summarized. 
\end{abstract}

\section{Introduction}
\label{intro}

Nearly forty years after Hawking's momentous discovery of black hole radiation, 
this parallel session testified to the continuing strength and increasing diversity of a field which still derives much inspiration from his insight. 
Indeed, the investigation of thermodynamic properties of black holes and causal horizons 
remains a vibrant line of research and was strongly represented in the session. 
Recent years have also seen many important developments in the theory of quantum fields
in curved spacetimes (QFT in CST). For example, the rigorous algebraic approach has achieved
many goals, including the construction of perturbative interacting theories, and deep
insights into the theory of general quantum fields through the formalism of local
covariance; ideas which were represented in a number of talks. In other directions, the session reported novel results in the application of QFT in CST to inflationary scenarios both through the study of backreaction (partly drawing on the insights of the algebraic approach) 
and also the special properties of QFT in de Sitter spacetime. Studies of the backreaction
naturally focus attention on the quantum stress-energy tensor and its properties, 
for example its violation of classical energy conditions and their replacement by
quantum inequalities. 

Hawking radiation, and its robustness relative to the ultrashort distance structure of spacetime,
have stimulated research in analogue models of gravity. These are typically condensed matter
systems in which perturbations propagate according to equations
mimicking those of a field theory on a curved background. Most interestingly, they show 
modified dispersion relations at high energies due to the intrinsically emergent nature of
the effective spacetime in the description. The session presented investigations both
of the effect of these modified dispersion relations on Hawking radiation as well as
of the lessons that can be drawn from these systems as toy models for an emergent spacetime or
quantum gravity.

Analogue gravity has also been an important stimulus for the investigation of
potential observational signatures of quantum gravity scanarios at low energy (below the Planck scale).
This field of research has rapidly developed from testing modified dispersion relations of the sort 
discussed above to the consideration of other possible effects and of viable frameworks that may
be tested in the future. Several talks on such quantum gravity phenomenology were presented, spanning
a variety of viewpoints. 

Finally, the issue of the backreaction of quantum effects on the geometry has motivated research into modifications
of Einstein gravity, via semi-classical and stochastic methods. Both were represented in the session.

In our view, the parallel session was very successful in representing a diversity of styles of research and
breadth of physical problems. It was one of the largest parallel sessions, with 33 talks and around the same
number of posters. Attendance was high, and lively discussion followed most of the talks. The talks themselves were of a
high quality and in many cases excellent. Indeed, two were selected for
prizes: Alexandre Le Tiec received one of the four Chandrasekhar Prizes for postdoctoral presentations,
while Valentina Baccetti received one of the ten Hartle Prizes for graduate student presentations. 

The variety of talks defies easy classification; in the following we have tried to group talks 
loosely by theme, but in reality many talks spanned more than one topic.

\section{Quantum field theory in curved spacetime}
\label{sec:QFTCST}

As mentioned above, the algebraic approach to quantum field theory in curved spacetimes has yielded many important developments in recent years. One impetus for this was the explicit formulation of local covariance
using methods from category theory~\cite{BrFrVe03}. Some models, like the 
free scalar field, fit easily into this framework. In the case of models
with gauge symmetry, however, the situation is more complex because
they can be sensitive to global topology. These issues were discussed by Dappiaggi, describing his recent work~\cite{Dappiaggi2} (with Sanders and Hack) and \cite{Dappiaggi1} (with Benini and Schenkel).  Maxwell theory, regarded as a theory of connections, was carefully quantized in the algebraic approach, emphasizing the topological aspects. Strict local
covariance fails, because certain observables can be  ``lost" when passing from one spacetime with nontrivial topology to a larger spacetime with trivial topology, for example.

Hollands' talk had different concerns: namely, the extent to which supersymmetry can be realized in QFT on specific spacetimes. The supersymmetries in question constitute the
conformal Killing superalgebra of a given fixed spacetime, on which
one then formulates theories based on SUSY-invariant Lagrangians. In a flat spacetime, the question of whether SUSY is realized at the quantum level amounts to whether the supersymmetries are implemented by operators commuting with the $S$-matrix; this approach is not
viable in curved spacetimes, owing to the lack of (or pathologies in) a global $S$-matrix.  Describing his work with de Medeiros~\cite{Hollands1}, Hollands explained
how the renormalization techniques developed for curved spacetime could be combined with BRST and Batalin-Vilkovisky formalisms to answer
this question at the perturbative level for $\mathcal{N}=2$ theories, 
in terms of whether or not the $\beta$-function of the theory vanishes identically, or whether the twistor spinors and conformal Killing vectors in the superalgebra obey certain conditions.

The Hawking effect has long been an inspiration for work in this area.  Sanders' talk centred on the existence of the Hartle--Hawking--Israel (HHI) state for a massive free scalar field on static black hole spacetimes. Existence outside the horizon was
established long ago by Sewell~\cite{Sewell1982} in the stationary case (and even for interacting theories); the question here was whether such a state extends across the horizon while maintaining the Hadamard form. 
Sanders answered this question affirmatively, by a careful analytic continuation argument to perform Wick rotation from the Euclidean case. A similar argument appears in an argument of Jacobson~\cite{Jacobson1994}, but uses continuation with respect to Killing time. Instead, Sanders employs a Gaussian normal coordinate, thus
avoiding the pathologies of Killing time at the horizon, and providing
a complete argument~\cite{Sanders1}. 

A different viewpoint on Hawking radiation was discussed by Pinamonti, who reported on recent papers with Moretti and Collini~\cite{Pinamonti1,Pinamonti2} on the Parikh--Wilczek derivation of the Hawking temperature based on  tunneling probabilities~\cite{PW2000}. 
Working in an algebraic approach, but {\em without} assuming a specific
field equation, Pinamonti described how the Parikh--Wilczek argument can be formalized and made rigorous, using only local arguments and in
a manner which extends to certain interacting theories. This resolves
a number of questions about the original derivation, such as its dependence on the specific geometry, quantum field state, and the 
regularization methods used. 

Moving mirror models have long been studied, as solvable models
in which particle creation and other quantum effects, such as negative energy densities, occur. Louko's talk, based on his papers with 
Bruschi, Faccio and Fuentes~\cite{Louko1} and Friis and Lee~\cite{Louko2}, explored QFT in an accelerated cavity with mirrored walls, with a view to applications in relativistic quantum information theory and potentially observable effects. By means of a systematic perturbative expansion for small accelerations, it was shown that resonances for
particle creation and mode mixing (a new effect) are obtained in periodic motion in $3+1$ dimensions; Louko argued that the mode mixing resonance is on the edge of current detection technology using desktop devices. Good's talk concerned the time-resolution of particle creation by massless scalar fields in the presence of moving mirrors in $1+1$-dimensions, based on work with Anderson and Evans~\cite{Good1}. A number of classes of mirror trajectory were discussed, for which full 
analytic descriptions of the particle production can be obtained. The time resolution of particle production is discussed by considering wavepacket states with suitable time and frequency localization. 

Several talks concerned the quantized stress-energy tensor and its properties. As is well-known by now, the stress-energy tensor violates the pointwise energy conditions of classical GR, and the energy density at a single point is unbounded below as the quantum state varies. In many models, quantum (energy) inequalities have been derived which show that the weighted energy density, obtained by integration along a smooth timelike curve with a smooth weight, is bounded from below in a state-independent way. 
These bounds even apply in curved spacetimes in considerable generality. The idea explored by Roman, reporting on work with Ford~\cite{Roman1}, was to consider the (unweighted) integral of the energy
density along portions of accelerated observer trajectories. It turns out the the integral can become
unboundedly negative as the proper time along the trajectory increases -- for example along an oscillatory trajectory with the quantum field in a suitable squeezed state. As Roman emphasized, these results do not contradict the quantum energy inequalities or vitiate their consequences, but rather demonstrate that the lower bounds can become quite weak over extended periods on accelerated trajectories. This might have consequences, for example, in considering potential defocussing effects on accelerated worldlines. 

In their talks, Ford and Bates considered the fluctuations of the stress-energy tensor, an issue that also played a role in Siemssen's presentation described in Sect.~\ref{sect:dS}. Ford presented results obtained with Roman and Fewster~\cite{Ford1,Ford2} concerning the probability distribution for individual measurements of the averaged energy density
in the vacuum state of Minkowski spacetime. The mean of this distribution vanishes, and its support
is bounded below (because of the quantum inequalities), so it is highly skewed and non-Gaussian. 
In $1+1$-dimensional conformal field theory, Ford reported a closed-form expression for the
probability distribution, assuming a Gaussian weight function~\cite{Ford1}. In $3+1$-dimensions, a closed form expression is not available, but a certain amount is known about the behaviour of the moments of the energy density e.g., for massless scalar and electromagnetic fields, and implies that the probability distribution has a slowly decaying tail at large positive values~\cite{Ford2}. Ford argued from this that large positive
fluctuations would dominate over thermal fluctuations at large energies, with potential implications
for rate estimates of black hole nucleation from the vacuum. Bates, describing work in progress, considered how the non-Gaussianity 
of such fluctuations might be used in stochastic gravity models; the linkage with the work of Ford {\em et al} is not entirely clear, as the fluctuation probability distributions differ. 

Another way in which classical and quantum dynamics can differ relates to their behaviour at boundaries or singularities. While classical equations
always require specification of boundary conditions at such points, 
quantum dynamics sometimes leaves no freedom to specify boundary conditions, thus resolving the singularity in a certain sense. This
was the subject of Konkowski's talk, based on her recent papers with
Helliwell \cite{Konkowski1,Konkowski2}. Following ideas of Horowitz and Marolf~\cite{HorMar1995}, a static spacetime is regarded
as being quantum mechanically nonsingular (resp., singular)  if the
spatial part of the Klein--Gordon operator is essentially self-adjoint (resp., fails to be essentially self-adjoint) on the domain of test functions supported away from the singularities, within the square-integrable functions on the spatial section.  Konkowski extended these ideas to   conformally static spacetimes and showed how some (but not all) singularities can be resolved in this framework. As she noted, it remains to be understood whether there is a unifying understanding of these phenomena. 

Kalinichenko reported on his computation of one-loop free energies
of scalar fields at high temperature (with Kazinski)~\cite{Kalinichenko1}. 
Making use of a heat-kernel expansion and by extensive calculations, 
the high-temperature expansion is obtained. The leading order term
represents Planck's law, to which the subdominant terms provide corrections. The final results are explicit in terms of heat-kernel coefficients and other geometric quantities.  Meanwhile, Groh also presented one-loop computations; in his case, for a class of 
``pure-connection'' theories that includes Krasnov's reformulation
of GR, and which can all be thought of as theories
of interacting gravitons. Groh's work, conducted with Krasnov and Steinwachs~\cite{Groh1}, has a long term aim of understanding whether this class of theories is closed under the renormalization group flow.
The computations reported by Groh, at one-loop about instanton backgrounds, represent a first step: the results differ from those of
metric GR, but perhaps less than might be expected, given that classical
equivalence is no guarantee of quantum equivalence.  

\section{De Sitter spacetime, inflation and semi-classical\\ gravity}
\label{sect:dS}

Quantum field theory in de Sitter spacetime has had a long and
sometimes controversial history, and several talks focussed
on aspects of the theory. Marolf, discussing recent work 
with Morrison and Srednicki \cite{Marolf1,Marolf2}, 
gave a perturbative construction of the $S$-matrix for
interacting scalar fields on global de Sitter spacetime. 
This is based on a discussion of the analytic structure
of such theories, for example, analogues of the K\"allen--Lehmann 
representation and detailed analysis of the pole structure. 
The resulting $S$-matrix has good properties, such as unitarity, invariance under field redefinition and the correct flat-spacetime limit. The $S$-matrix
is not experimentally accessible to a single observer, but Marolf 
emphasized its importance as a theoretical tool and a bridge 
between QFT in Minkowski space and de Sitter; for example,  unitarity
gives rise to a de Sitter optical theorem. This theme was
developed further by Morrison, who focussed
on a non-perturbative $S$-matrices for principal chiral models in $1+1$-dimensional de Sitter
spacetime. These are nonlinear $\sigma$-models whose classical
conformal invariance is broken under quantization. However, Morrison described how the
$S$-matrix can be computed by bootstrap techniques and consideration
of the asymptotic states and conserved charges. 

Woodard's talk, based on work with Degueldre~\cite{Woodard1} and Prokopec and Tornkvist~\cite{Woodard2}, concerned the effects of vacuum polarization in an inflating universe. It has been proposed that vacuum polarization of charged scalars during inflation gives rise to an effective photon mass and hence to `quantum corrected' electromagnetism. The issue discussed in Woodard's talk was what
effect this has on the electric and magnetic fields of point sources, 
fixed at the spatial origin in conformal coordinates, as seen in the
frame of co-moving or static observers. These observers see
the effects of vacuum polarization as providing either a screening (conformal observer)
or enhancement (static observer) of the classical result. These modifications to the standard result become nonperturbative after only a few $e$-foldings.

Two talks concerned the issue of quantum fluctuations in an inflationary
scenario. Siemssen reported his work with Pinamonti~\cite{Siemssen1} on the semi-classical Einstein equation. The idea, somewhat as in stochastic gravity, is to take the Einstein tensor as a random field, with its moments
equated to those of the renormalized stress-energy tensor of a 
quantum field -- here taken as the conformally coupled massive scalar field in the Bunch--Davies state. This is used to compute curvature fluctuations in a perturbed inflating universe, in a model for which
only the trace of the stress-energy tensor acts as a source, and
leading to an `almost' scale-invariant power spectrum of the fluctuations.  
Marozzi considered more general fluctuations about a spatially flat FLRW background geometry in a chaotic inflation model. In particular, 
the idea is to consider observables (and observers) associated with
test fields in a gauge invariant formalism. Based on papers~\cite{Marozzi1,Marozzi2}, Marozzi described how effective expansion rates and equation of state could be computed for such observers, allowing the effect of long-wavelength fluctuations of the background to become manifest.

\section{Black holes}

Black hole evaporation and the associated thermodynamical aspects of black mechanics have been for many years a driving force for the main subject of this parallel session. The talks present offered quite a representative overview of the main discussions still active in the very broad community working on the field. In the particular case of the possible information loss in black hole evaporation and the so called firewall hypothesis we chose to co-organize, together with the chairs of sessions D1 and D2, a separate parallel session (also reported in these proceedings). Of course, some talks of D4 still had some connection with the above mentioned topics. 

This is the case for the talk given by Smerlak, reporting results mainly presented in \cite{Smerlak1,Smerlak2}. In fact, the discussion focussed here on the possible outcome of detector measurements for different types of observers, particularly those free falling at the horizon along more or less bounded orbits. In contrast to standard accounts, it was claimed that freely falling observers following bounded orbits (energy per unit mass along the geodesic $E$ less than one) would indeed detect an Hawking flux even at horizon crossing. Actually, for highly bounded trajectories, Hawking radiation appears to be dominated by ingoing modes and becoming arbitrarily hot as $E\to 0$. These results are not directly related to the firewall proposal but seem to resemble some of its conclusions in a purely QFT in curved spacetime approach.

As said Hawking radiation corroborated and made consistent the thermodynamical properties of black hole mechanics. This remarkable connection between thermodynamics and gravity was further explored in the talk prepared by Haggard and presented by his co-author Rovelli \cite{Haggard1} (in Haggard's absence, due to illness). The zeroth law of thermodynamics, stating that temperature is uniform at thermal equilibrium, is notoriously violated in relativistic gravity. Temperature uniformity is often derived from the maximisation of the total number of microstates of two interacting systems under energy exchanges. The talk presented a generalised version of this derivation, based on informational notions, which remains valid in a general relativistic context. The results suggest a new principle for equilibrium that might be used to progress towards the  foundation of a general covariant statistical mechanics.

The extension of black hole thermodynamics beyond stationarity and axisymmetry was instead the subject of investigation presented in the talk by  Le Tiec (derived from a paper with Gralla \cite{GrallaLeTiec1}). In particular, the seminar discussed the zeroth and first law of black hole mechanics in the setting of a rotating black hole perturbed by a ``corotating moon'' i.e.~one with angular velocity equal to that of the unperturbed event horizon. For the zeroth law, helical symmetry of the spacetime and corotation imply the vanishing expansion and shear of the perturbed future event horizon. Then rigidity theorems can be used to prove that this has to be a Killing horizon, which must have then constant surface gravity.
With regard to the first law, the analysis showed how the effect of perturbations preserves the entropy of the black hole but reduces the temperature (which is still proportional to the surface gravity) by a negative constant amount, implying that the black hole is effectively cooled down by the presence of the corotating moon.

The very nature and meaning of entropy was instead the main subject of the talk by Baccetti (from a paper in collaboration with Visser \cite{Baccetti1}) who presented an investigation aimed at generalising the so called spacetime thermodynamics approach started by Jacobson in the nineties \cite{Jacobson_for_Baccetti}. The work focussed on the notion of
Clausius entropy associated to the matter crossing arbitrary bifurcate null surfaces in arbitrary spacetimes. The main result implied that one can associate an observer-dependent notion of entropy, very closely related to the Clausius entropy, and a generalisation of Jacobson's local-Rindler entropy, to any arbitrary bifurcate null surface. Some considerations for the impact of such study on the spacetime thermodynamics proposal were also discussed.

The concept of entropy and its extension to black holes was also the main interest of Quinta's talk (from work in progress with Lemos and Zaslavskii). Extremal black holes have been  a subject of debate for many years now. These are the zero temperature objects in black hole thermodynamics, and while they seem to satisfy the unattainability formulation of the third law they also appear to violate the so called isoentropic formulation of that principle: if a Bekenstein--Hawking entropy can be associated to extremal black holes then their entropy would not be universal (i.e.~independent of the macroscopic properties of the black hole such as mass, charge or angular momentum). Quinta's talk focussed on the entropy of extremal black holes by considering the limit case of the collapse of thin charged shells. The results suggest  that the extremal case remains basically unconstrained, hinting at a rather different behaviour from the non-extremal case. The author suggests this might be taken as a sign that that extremal black holes have an undetermined entropy that does not need to be equal to the Bekenstein--Hawking one.

The talk by Winstanley focussed instead on the quantization of a massless fermion field on a non-extremal rotating black hole~\cite{Winstanley1}. Since all fermion modes (both particle and anti-particle) have positive norm, they offer much greater flexibility in how quantum states are defined compared with the bosonic case. In particular, it was shown that fermions, at odds with the scalar field case, do allow for a Kerr black hole to define two quantum states which are analogues of the standard Boulware state (i.e~empty at both past and future null infinity) and Hartle--Hawking one (thermal bath of fermions surrounding
the black hole).

Of course quantum fields do back scatter on curved spacetimes,
inducing  greybody factors that are very important in determining the equilibrium properties on black hole spacetimes. The talk by  Crispino~\cite{Oliveira:2011zz} presented a study aimed at determining the electromagnetic and gravitational absorption cross sections on an extreme Reissner-Nordstr\"om black hole spacetime (a maximally charged non-rotating black hole). After reviewing the absorption for the  Schwarzschild spacetime case, the discussion focussed on this class of solutions showing an equality between gravitational and electromagnetic absorption cross sections for all frequencies. Consequently extremal charged black holes are the first example, in general relativity, in four dimensions, of an object that absorb the electromagnetic and gravitational waves in exactly the same way.

The talk by  Bambi~\cite{Bambi1} focussed on the possibility of distinguishing black holes from naked singularities. The cosmic censorship conjecture --- according to which singularities produced in the gravitational collapse cannot be seen by distant observers and must be hidden within black holes --- remains an important open question. The work presented, discussed the possibility of observationally testing the conjecture by studying the radiation emitted by a collapsing cloud of dust as a possible probe to distinguish the birth of a black hole from that of a naked singularity. In the simple dust model considered, it is found that the properties of the radiation emitted in the two scenarios are unfortunately qualitatively similar. This seems to imply that observational tests of the cosmic censorship conjecture may be very difficult to perform in a near future (or possibly even in principle impossible).

\section{Analogue gravity and Quantum gravity phenomen\-ology}

Black holes and their evaporation have stimulated research beyond the standard QFT in curved spacetime. In particular a notable approach that has gained momentum in the last 30 years is the study of so called Analogue Gravity. Since the seminal paper by Unruh \cite{Unruh:1980cg} that noticed the analogy between the propagation of linearised perturbations on a perfect fluid flow and the dynamics of scalar fields on a curved background, a stream of research has developed aimed at using these condensed matter systems for testing QFT in curved spacetimes (such as Hawking radiation or cosmological particle creation) as well as ideas about the possible microscopic/quantum description of gravity.

The talk by Coutant summarised  present knowledge of the description of Hawking radiation in analogue system. These are characterised by the presence of modified dispersion relations for the linearised perturbations playing the role of the propagating matter fields. Such modifications are associated to the effects at high energy/short wavelengths of the underlying microscopic structure of the emergent spacetime provided by the fluid flow and are associated to a breakdown of the acoustic Lorentz invariance. The seminar discussed the robustness of Hawking radiation to such ultraviolet modifications (a very cogent problem given the so called trans-Planckian problem in black hole physics) and showed the relevant physics at the base of the particle creation around the acoustic horizon (the locus of points where the flow speed matches the local speed of sound which is the analogue of the event horizon on black holes).   Among other things it was shown that the acoustic horizon is effectively broadened (acquires an effective thickness) as a consequence of dispersion. Finally, Coutant gave a novel description of an instability associated to the excitation of a zero energy (but high momentum) mode at white hole horizons, the so called undulation. Such effect could have potentially interesting observational applications.

Another talk on analogue gravity models was given by Carballo-Rubio. In this case the focus was on another growing trend of studies in Analogue Gravity which is aimed at using these systems as toy models of emergent gravitational dynamics, not just of an emergent spacetime and of a QFT above it. Indeed, while analogue gravity has shown quite vividly that it is not difficult to obtain the latter, it has proven the point that gravitation-like dynamical equations (which are related to a local gauge invariance) are quite more difficult to obtain. The work presented by Carballo-Rubio (in progress with Barcel\'o, Garay and Jannes) approaches this problem by studying a simpler situation, namely, the emergence of electrodynamics in a $^3$He-like system. The work presented, considered a non-relativistic system composed by two families of massive, interacting, fermions and described its superfluid phase. In this phase, and with an homogeneous order parameter, quasiparticle excitations were shown to propagate as fermion fields on a flat spacetime. The gist of the proposal is that in the case of an inhomogeneous order parameter --- a priori a simple non-dynamical, background gauge field --- the quasi-particle excitations can produce an emergent dynamics for such field in the same fashion as vacuum fluctuations of matter fields on a curved background can produce a gravitational action in Sakharov's induced gravity scenarios. Preliminary results in this sense were presented.

Analogue models of gravity have recently inspired a series of works aimed at testing possible signatures of new physics associated with the microscopic nature of spacetime at the Planck scale. This stream of research goes under the broad denomination of Quantum Gravity phenomenology (QGPh). As mentioned above, Lorentz invariance might be violated at the Planck scale, however there is an alternative point of view that a new relativity group might be realised in which the UV modified dispersion relations are invariant under a new set of transformations and momenta are composed non linearly. This idea goes under the broad name of doubly or deformed special relativity (DSR).  A way of testing both Lorentz breaking and Lorentz deformed scenarios consists of using purely kinematical effects such as the differences in time of flight for elementary particles at different energies induced generically by the above mentioned modified dispersion relations. 

The talk by Amelino-Camelia discussed explicitly two promising developments in this sense consisting of the recent detection of high energy neutrinos by the IceCube Collaboration and of photons from energetic gamma ray bursts by FERMI LAT (see \cite{Amelino-Camelia:2013jga,Amelino-Camelia:2013naa}. In the case of neutrino detection it was shown that some of the detected neutrinos could be correlated with GRB emission within a quantum spacetime dispersive scenario, while for the GRB detection a comparison of the observed features for GRBs at different redshifts provides some encouragement for a redshift dependence of the effects of the type expected for a quantum-spacetime interpretation.  It was however stressed that other aspects of the analysis appear to support instead an interpretation of the above mentioned effects based on intrinsic properties of GRBs, and that further data are needed in order to discriminate among alternative explanations.

A recent incarnation of the above mentioned DSR scenarios consists of the proposal that momentum space is fundamental and curved while coordinate space has to be locally determined as the tangent space to it. In brief, in momentum space a metric is determined by the dispersion relations and an independent connection is fixed by the non-linear composition rule.  Within such a framework the notion of locality (a coincidence of events) becomes relative (i.e.~observer dependent), hence the name of Relative Locality.  The talk by Kowalski-Glikman dealt with two issues related to this framework (see e.g.~\cite{KowalskiGlikman1,KowalskiGlikman2}). The first part of the talk  discussed the so-called soccer ball problem, that is, the apparent incompatibility between DSR scenarios and macroscopic (heavier than Planck mass) objects. The point made was that by  formalising a macroscopic object  as a rigid system of $N$ constituents, each having exactly the same momentum (and modified dispersion relations), one may see that the effect of the deformation is suppressed by $N$ times the deformation scale, hence allowing a classical behaviour for large bodies.  The second part of the talk focussed on a recent objection to the framework based on the fact that unacceptably large amplitudes seem to be predicted for the momentum fluctuations of macroscopic
bodies at finite temperature~\cite{Hossenfelder:2012vk}. The main claim was that, by suitably taking into account contributions from the geometry of momentum space in the calculation, the framework does admit some specific geometries for which the above mentioned amplitudes are indeed small.

The session offered other two talks about Relative Locality. The one by  Banburski (see \cite{Banburski:2013jfa}) discussed the possibility 
of having a non local deformation that preserves full Lorentz invariance
in 3+1 dimensions. It was shown how to construct a homogeneously curved momentum space preserving the full action of the Lorentz group in four dimensions and higher, relaxing locality. This momentum space has the geometry of de Sitter space. Finally, the properties of this relative locality scenario were discussed and it was shown that they imply a non-commutativity of the Snyder type.

An alternative take on relative locality scenarios was presented in the talk by Chen in which implications for causality were discussed (see \cite{Chen1,Chen2}). While in General Relativity the causal relationships between events are expressed as geometrical relationships between points on the spacetime manifold, in Relative Locality there is instead no universal spacetime manifold (which is rather a derived concept from the curved momentum one). The analysis considered causal loops, an example being a particle, say $0$, created at a spacetime event A and destroyed at some other event B joined with a particle, say $1$ created at B and destroyed at A so to generate particle $0$ there. Such causal loops are normally forbidden in Special Relativity. A detailed consideration of the orientability properties of such loops seems to lead to the conclusion that for Relative Locality causality violations can show up at least for some specific choices of points on the cotangent space.

Quantum gravity phenomenology is also an experimental/observational endeavour and the last two talks of our summary exactly dealt with these efforts.  Many quantum gravity scenarios has suggested that quantum gravity might introduce an extra uncertainty on the localisation of objects given that the use of high energy probes close to Planck-scale distances end up necessarily perturbing the local spacetime geometry. This observation has motivated the proposal of the so called generalised uncertainty relations, which imply changes in the energy spectrum of quantum systems. As a consequence, quantum gravitational effects could be revealed by experiments able to test deviations from standard quantum mechanics such as those recently proposed on macroscopic mechanical oscillators. In the work presented,  Cerdonio presented the results of an experiment based on the resonant gravitational bar AURIGA~\cite{Cerdonio1}. These consisted in the exploitation of the sub-millikelvin cooling of the normal modes of the ton-scale gravitational wave detector which allowed to place a new strong upper limit for possible Planck-scale modifications on the ground-state energy of the oscillator. 

Another, more traditional, arena for possible observational effects of quantum gravity is of course quantum cosmology. The talk by Kr\"amer~\cite{Kraemer1,Kraemer2} discussed a derivation of the primordial power spectrum of density fluctuations within such a framework of quantum cosmology. This was obtained by performing a Born--Oppenheimer approximation to the Wheeler--DeWitt equation for an inflationary universe with a scalar field. After recovering the scale-invariant power spectrum that is found as an approximation in the simplest inflationary models, the quantum gravitational corrections to this spectrum were derived. Unfortunately, it appears that no measurable signatures in the CMB anisotropy spectrum will be seen in the near future due to the cosmic variance (i.e., the statistical uncertainty inherent in observations of the universe at the largest scales). Nonetheless, the non-observation so far of such corrections translates into an upper bound on the energy scale of inflation.

In conclusion, we had a very lively parallel session with a very broad range of topics.  We look forward to a new D4 session in GR21.

\end{document}